\documentclass[times,10pt,onecolumn]{article}

\usepackage{times,amsmath,epsfig}
\usepackage{isomath}
\usepackage[ruled]{algorithm2e}
\usepackage{graphicx}
\usepackage{amssymb}
\usepackage{latexsym}
\usepackage{subfigure}
\usepackage{epstopdf}
\usepackage{amsmath}
\usepackage{mathtools}
\usepackage[belowskip=-13pt,aboveskip=2pt]{caption}

\def\qed{\ifmmode\squareforqed\else{\unskip\nobreak\hfil
    \penalty50\hskip1em\null\nobreak\hfil\squareforqed
    \parfillskip=0pt\finalhyphendemerits=0\endgraf}\fi}

\def\triaforqsd{\hbox{\Large$\triangleleft$}}
\def\qsd{\ifmmode\triaforqsd\else{\unskip\nobreak\hfil
    \penalty50\hskip1em\null\nobreak\hfil\triaforqsd
    \parfillskip=0pt\finalhyphendemerits=0\endgraf}\fi}

 \newcommand{\A}{\mathcal{A}}

\newcommand{\ti}{v}

\newcommand{\D}{\mathcal{D}}

\newcommand{\T}{\mathcal{T}}
\newcommand{\U}{\mathcal{U}}

\newcommand{\LP}{\mbox{\texttt{LP}}}
\newcommand{\G}{\mathcal{G}}
\newcommand{\I}{\mathcal{I}}

\newcommand{\NDUP}{\texttt{NDUP}}
\newcommand{\NTRANS}{\texttt{NTRANS}}
\newcommand{\msIFM}{\mbox{$ms\texttt{-}\mathtt{IFM}$}}

\newcommand{\IFMI}{\mbox{$\mathtt{IFM}_\mathtt{I}$}}
\newcommand{\IFMG}{\mbox{$\mathtt{IFM}_\mathtt{G}$}}
\newcommand{\IFM}{\mbox{$\mathtt{IFM}$}}

\newcommand{\FREQSAT}{\mbox{$\mathtt{FREQSAT}$}}

\newcommand{\NP}{\mbox{$\mathtt{NP}$}}

\newcommand{\nexp}{\mbox{$\mathtt{NEXP}$}}
\newcommand{\pspace}{\mbox{$\mathtt{PSPACE}$}}
\newcommand{\PP}{\mbox{$\mathtt{PP}$}}

\newcommand{\sat}{{\varsigma}}
\newcommand{\size}{\mbox{$\mathtt{size}$}}

\newtheorem{example}{Example}
\newtheorem{definition}{Definition}
\newtheorem{proposition}{Proposition}
\title{Multi-Sorted Inverse Frequent Itemsets Mining}

\author{%
{Domenico Sacc\`{a}{\small $~^{1}$}, Edoardo Serra{\small $~^{2}$}, Pietro Dicosta{\small $~^{1}$}, Antonio Piccolo{\small $~^{1}$} }\\ \\%
\small
$^{1}$\,DIMES Department, University of Calabria, {\em \{sacca, piccolo, dicosta\}@unical.it} \\ 
\small
$^{2}$\,Computer Science Department, University of Maryland, {\em eserra@umiacs.umd.edu}\\
}

\date{}
\begin{document}

\maketitle

\pagestyle{plain}

\begin{abstract}
The development of novel platforms and techniques for emerging ``Big Data'' applications requires the availability of real-life datasets for data-driven experiments, which are however  out of reach for academic research in most cases as they are typically proprietary.
A possible solution is to use synthesized datasets that reflect patterns of real ones 
in order to ensure high quality experimental findings.
A first step in this direction is to use inverse mining techniques such as  inverse frequent itemset mining (\IFM) that consists of generating a transactional database satisfying given support constraints on the itemsets in an input set, that are typically the frequent ones. 
This paper introduces an extension of \IFM\, called {\em many-sorted \IFM}, where the schemes for the datasets to be generated are those typical of  Big Tables as required in emerging big data applications, e.g., social network analytics.

\end{abstract}

\pagenumbering{arabic}

%\IEEEpeerreviewmaketitle

%\category{F.2}{Theory of Computation }{Analysis of Algorithms and Problem Complexity}
%A category including the fourth, optional field follows...
%\category{H.2.8}{Information Systems}{Database Management}[Database Applications: data mining]
%\category{H.2.2}{Information Systems}{Database Management}{Logical Design}

%\terms{Theory, Design}

%\keywords{Inverse data mining, Frequent Itemsets, NEXP Completeness, Clumn generation simplex} % NOT required for Proceedings

%\noindent \textbf{Keywords:} frequent pattern mining.
% ----------------------------------------------------------------
\section{Introduction}

Emerging ``Big Data'' platforms and applications call for the invention of  novel data analysis techniques that are capable to handle large amount of data \cite{MC.2013.196}. There is therefore an increasing need to use real-life datasets for data-driven experiments but, as pointed out in a recent ACM SIGMOD Blog post by
Gerhard Weikum \cite{BlogWeikum13}, datasets used into research papers are often poor. Companies have their own interesting data, and industrial labs have access to such data and real-life workloads; however, such datasets are often proprietary and out of reach for academic research.
In order to ensure high quality experimental findings, inverse mining techniques can be applied to generate artificial datasets that reflect the patterns of real ones: the patterns are first discovered by data mining techniques (or even directly provided by domain experts) and then used to generate ``realistic'' privacy-preserving datasets.

In order to enlarge the application domain of \IFM, we introduce a further extension that considers more structured schemes for the datasets to be generated, as required in emerging big data applications, e.g., social network analytics. We assume that the set $\I$ of items is partitioned into $1+p$ classes: $\G$ with $n$ items ({\em group items}) and $A_1, \dots, \A_p$ with respectively $n_1, \dots, n_p$ items ({\em  single items}). A {\em many-sorted transaction} $I$ is a set of items $\{a_1, \dots, a_p \} \cup G$, where $a_1 \in A_1, \dots, a_p \in A_p$ and $G \subseteq \G$, i.e., $I$ consists of a classical itemset of $\G$ extended with exactly one item $a_i$ for every set $A_i$ of items. As an example of a many-sorted dataset, consider a social network application with members  characterized by the attributes   {\em Gender}, {\em Location} and {\em Age}.  The domains of these attributes are sets of single items.  A member may belong to various groups, whose values are stored into the set $Group$ of group items. A many-sorted transaction such as $\{Male, Rome, 25, g_1, g_4\}$ represents a 25-year old male member located in Rome who belongs to the groups $g_1$ and $g_4$. Note that, as the attributes do not define a key, there may exist several occurrences of the same member, i.e., a many-sorted transaction actually represents a uniform group of members. 

We define a many-sorted extension of \IFM\ called $ms$-\IFM, for which not all itemsets can be transactions, e.g., any transaction must have exactly one item for the attributes  {\em Gender}, {\em Location} and {\em Age}. In this framework, duplicate constraints may have an important role to define patterns to be incorporated in the generated datasets. Two important results are: (1) the complexity of $ms$-\IFM\ is the same as for classical \IFM\ and (2) the extended column generation algorithm can be easily adapted to solve  $ms$-\IFM\ by means of a suitable representations of the variables associated to transaction occurrences.

The remainder of the paper is organized as follows. Section \ref{prelim-section} introduces basic notation and illustrates  results from  recent literature on \IFM. We extend the \IFM\ problem to the domain of big data application in Section \ref{sect:msIFM} and define the $ms$-\IFM\ problem in Section \ref{sect:msIFM:problem}. Later on, we formulate the  $ms$-\IFM\ problem as a succinct linear program in Section \ref{sub:ILP} and present an extension of the column-generation simplex for its resolution in Section \ref{sect:cg:simplex}. Finally we draw the conclusion and discuss further work in Section \ref{sect:concl}.

\section{Preliminaries and Related Work}\label{prelim-section}

Let $\I$ be a finite domain of $n$ elements, also called \emph{items}. Any subset $I \subseteq \I$ is an \emph{itemset} over $\I$. 
A ({\em transactional}) {\em database} $\D$ over $\I$ (also called {\em dataset}) is a bag of itemsets, which may occur duplicated in $\D$ --- the size $|\D|$ of $\D$ is the total number of its itemsets, called \emph{transactions}. 

Given a database $\D \subseteq \I$, for each  itemset $I \in D$, there exist two important measures: (i)
the \emph{number of duplicates} of $I$, denoted as $\delta^{\D} (I)$, that is the number of occurrences of $I$ in $\D$, and 
(ii)  the \emph{support} of $I$, denoted as $\sigma^{\D} (I)$, that is the sum of all number of duplicates of itemsets in $\D$ containing $I$, i.e., $\sigma^{\D} (I) = \sum_{J \in \D \wedge I \subseteq J} \delta^{\D} (J)$ -- an alternative measure is the frequency  $f^{\D} (I)=\sigma^{\D} (I) / |\D|$.
A database $D$ can be represented in a succinct format as a set of pairs $(I,\delta^{\D} (I))$. 

We say that $I$ is a \emph{frequent} (resp., {\em infrequent}) itemset in $\D$ if its support is greater than or equal to (resp.,  less than) a given threshold. 
A popular mining task over transaction databases is to single out the set of the \emph{frequent/infrequent itemsets} \cite{AgrawalIS93,GunopulosKMT97,HanCXY07,GoetZaki04}. 

The perspective of the frequent itemset mining problem can be naturally inverted as follows:  we are be given in advance a set of itemsets together with their frequency constraints and our goal is then to decide whether there is a transaction database satisfying the above constraints (and, of course, compute the database whenever the answer is positive). This problem, called the \emph{inverse
frequent itemset mining} problem ($\IFM$), has been introduced in the context of defining generators for benchmarks of mining
algorithms~\cite{Mielika03}, and has been subsequently reconsidered in privacy preserving contexts~\cite{PPDM,WuWWL05}).
$\IFM$ has been proved to be in $\pspace$ and $\NP$-hard. As discussed in Section 1, the original \IFM\ formulation does not introduce any constraint on infrequency.

A reformulation of \IFM\ in terms of frequencies instead of supports has been introduced in \cite{Calders04,Calders07} with the name $\FREQSAT \small{\{\NTRANS\}}$. The two  problems are equivalent and have been shown to be in $\pspace$ and $\NP$-hard. The basic version of the frequency formulation, called \FREQSAT, does not fix the number $\NTRANS$ of transaction in a feasible database -- the corresponding decision problem has been proved to be $\NP$-complete. A further variant of the problem has been introduced in \cite{Calders04,Calders07}
with the name $\FREQSAT \small{\{\NTRANS,\NDUP\}}$: all itemsets may occur as transactions in $D$ at most a fixed number of times ($\NDUP$). This problem is in $\pspace$ and $\PP$-hard.

A simple solution to exclude unexpected frequent itemset from a feasible solution is the formulation proposed in \cite{GuSaSe09}, which is called $\IFM_S$: only itemsets in $S$ can be included as transactions in $\D$. The decision complexity of this problem is $\NP$-complete as stated in \cite{GuSaSe09} and proved in the Appendix.
The version of \IFM\ with infrequency support constraint ($\IFMI$ for short), has been recently proposed in \cite{MocciaEtAl12} and its decision complexity is $\nexp$-complete as proven in \cite{SaccaSerra12}. 

\section{IFM for Big Data Applications}\label{sect:msIFM}

In this section we provide an extension of the inverse frequent itemsets mining problem for generating a dataset with a more elaborated schema: a {\em many-sorted dataset} is a big table on a NOSQL relation $R(K, A_1, \dots,  A_p, G_1,\dots,$ $G_q )$, where $K$ is the table key, $A_1, \dots,  A_p$ are classical single-valued (SV) attributes and
$G_1,\dots,G_q$ are multi-valued (MV) attributes. 
Let $ \A_1, \dots,  \A_p, \G_1, \dots,  \G_q$ be the finite domains respectively for the attributes $A_1, \dots,  A_p, G_1, \dots,$ $G_q$, where for each $i, 1 \leq i \leq p$, $|\A_i|=  \dot{n}_i$ and for each $i, 1 \leq i \leq q$, $|\G_i|=  \ddot{n}_i$  -- we assume that the values of these domains (called SV or MV {\em items}) are given in input and all domains are pairwise disjoint. On the other hand, the domain of the key $K$ is countably infinite and its values are not listed. 

We construct the set $\I$ of items as the union of all the domains $\A_1, \dots,$ $ \A_p, \G_1, \dots,$ $\G_q$. Then the set $\I$ of all items is partitioned into $p+q$ classes, one for each composing domain. Let $\dot{n} = \sum_{i=1}^{p} \dot{n}_i$ and $\ddot{n} = \sum_{i=1}^{q} \ddot{n}_i$; then $n = |\I| = \ddot{n}+\dot{n}$.

A {\em many-sorted transaction} $I$ is a set of items $\{a_1, \dots, a_p \} \cup J_1 \cup \dots \cup J_q$, where $a_1 \in \A_1, \dots, a_p \in \A_p$, $J_1 \subseteq \G_1, \dots$ and $J_q \subseteq \G_q$, i.e., $I$ consists of the union of $q$ classical (possibly empty) itemsets, one for each MV attribute, extended with exactly one item for every SV attribute. 
A {\em many-sorted itemset} $I$ is any (not necessarily proper) subset of a many-sorted transaction -- i.e., $I$ consists of a classical itemset of $\G$ extended with at most one item for every set of single items. A {\em many-sorted dataset} $\D$ is a set of pairs $(I,\delta^\D(I))$, where $I$ is a many-sorted transaction and $\delta^\D(I)$ is the number of occurrences of $I$ in $\D$. 
The size of $\D$ is $\delta^\D = \sum_{I \in \D} \delta^\D(I)$.
(In the following, we shall omit the term {\em many-sorted} whenever it is clear from the context.)

Let $\T_\I$ and $\U_\I$ be the sets of all transactions and of all itemsets, respectively. The cardinalities of $\T_\I$  and of $\U_\I$ are $2^{\ddot{n}} \cdot \prod_{i=1}^{p} \dot{n}_i$ and $2^{\ddot{n}} \cdot \prod_{i=1}^{p} (\dot{n}_i+1)$. 

Given $I \in \T_\I$ and any SV attribute $A_i$, $I_{A_i}$ denotes the value of  $\A_i$ in $I$. Similarly, given any MV attribute $G_i$, 
$I_{G_i}$ denotes the (possibly empty) set of values of  $\G_i$ in $I$.

A {\em SV selection} is a pair $(A_i, a)$, where $A_i$ is any SV attribute and $a$ is any value in $\A_i$. 
A {\em MV selection} is a triple $(G_j, J, *)$, where $G_j$ is any MV attribute, $J \subseteq \G_j$ and ``*'' is either ``$=$'' ({\em equality MV selection}) or ``$\subseteq$''  ({\em subset MV selection}). 

A {\em selection list} $L$ is a non-empty list of SV and MV selections such that there are no two distinct selections in $L$ with the same attribute. We say that $L$ is {\em full} if all attributes occur in it. Given a transaction $I$, we say that $\sat(L,I)$ is {\em true} if both for every SV selection $(A_i, a)$ in $L$, $I_{A_i}=a$ and for every MV selection $(G_j, J,*)$ in $L$, 
$J * I_{G_j}$.

Given a selection list $L$ and two integers $\sigma_1$ and $\sigma_2$ for which $0 \leq \sigma_1 \leq \sigma_2$, $\gamma_\sigma = \langle L, \sigma_1, \sigma_2\rangle$ represents a {\em support constraint} defined as follows. Given a database $\D$, $\D \models \gamma_\sigma$ (i.e., $\gamma_\sigma$ is satisfied by $\D$) if:
$$\sigma_1 \leq \ \ \  \sum_{\mathclap{I \in \D \wedge \sat(L,I)}} \  \  \delta^\D(I)  \leq \sigma_2.$$
Given a set $\Sigma$ of support constraints and a database $\D$, $\Sigma$ is satisfied by $\D$ ($\D \models \Sigma$), if for each $\gamma_\sigma \in \Sigma$, $\D \models \gamma_\sigma$.

We call $\langle L, \sigma_1, \sigma_2\rangle$ a {\em domain support constraint} if (i) both $L$ is a singleton and includes a SV or a subset MV selection or (ii) {\em many-sorted support constraint} otherwise.

Given a selection list $L$ and an integer $\delta_2 > 0$, $\gamma_\delta=\langle L, \delta_2\rangle$ represents a {\em duplicate constraint} defined as follows. Given a database $\D$, $\gamma_\delta$ is satisfied by $\D$ (written as $\D \models \gamma_\delta$) if for each $I \in \D$ for which 
$\sat(L,I)$ is true, 
$\delta^\D(I) \leq \delta_2.$
Given a set $\Delta$ of duplicate constraints and a database $\D$, $\Delta$ is satisfied by $\D$ ($\D \models \Delta$), if for each $\gamma_\delta \in \Delta$, $\D \models \gamma_\delta$.

\begin{example}
{\em Individuals are characterized by the SV attributes   {\em Gender}, {\em Location} and {\em Age} and by the MV attributes {\em Groups} and {\em Events}:  an individual may belong to various groups and may attend a number of events.
A  transaction $I=\{Male, Rome, 25, g_1, g_4,$ $e_1, e_3\}$ represents an individual  a 25-year old male individual located in Rome who belongs to the groups $g_1$ and $g_4$ and attends the events $e_1$ and $e_3$. Note that, as the attributes do not define a key, there may exist several occurrences of the same individual. The transaction $J=\{Female, Rome, 20, g_1, g_2\}$ represents an individual who does not attend any event. Examples of constraints are
\begin{itemize}
\item {\em Domain support constraints}: \\ $\langle[(Gender, Male)], 4,000,000, 6,000,000\rangle$ states that the number of male individuals in a feasible dataset must be in the range from 4 to 6 millions; \\ $\langle[(Groups, \{g_1,g_2\},\subseteq)], 100,000, 200,000\rangle$ states that the number of individuals in a feasible dataset who are participating to at least the groups $g_1$ and $g_2$ must be between 100,000 and 200,000, while \\ $\langle[(Groups, \{g_1,$ $g_2\},=)], 5000, 8000\rangle$ states that the number of individuals in a feasible dataset who are participating to exactly the groups $g_1$ and $g_2$ must be between 5000 and 8000;

\item {\em Support constraints}: \\ $\langle [(Gender, Male),(Location,Rome),$ $(Groups,$ $\{g_1,g_2\},$ $\subseteq)],$ $10000, 20000\rangle$ states that the number of male individuals in a feasible dataset who are located in Rome and are participating to at least the groups $g_1$ and $g_2$ must be in the range from 10000 to 20000; \\ $\langle[(Gender, Female),(Groups,\{g_1,$ $g_2\},\subseteq), (Events,\{e_1,e_3\},\subseteq)], 500, 1000\rangle$ states that the number of female individuals in a feasible dataset who are participating to at least the groups $g_1$ and $g_2$ and attending at the least the events $e_1$ and $e_3$ must be in the range from 500 to 1000; 

\item {\em Duplicate constraints}: \\ $\langle [(Gender, Male),(Location,Rome), (Groups,\{g_1,g_2\},$ $\subseteq)], 1500 \rangle$ states that the number of every male individual in a feasible dataset who is located in Rome and is participating to at least the groups $g_1$ and $g_2$ must be less than or equal to 1500; \\ $\langle[(Gender, Female),(Group,\{g_1,g_2\},\subseteq), (Events,$ $\{e_1,e_3\},\subseteq)],  2000\rangle$ states that the number of every female individual in a feasible dataset who is participating to at least the groups $g_1$ and $g_2$ and attending at the least the events $e_1$ and $e_3$ must be less than or equal to 2000. 
\hfill $\Box$
\end{itemize}
}
\end{example}

\section{Many-Sorted IFM Problem}\label{sect:msIFM:problem}

In this section we provide a general formulation of the many-sorted inverse frequent itemsets mining problem. 
Let $R(K, A_1, \dots,  A_p, G_1,\dots,$ $G_q )$ be a NOSQL relation, where $K$ is the table key, $A_1, \dots,  A_p$ are SV attributes and
$G_1,\dots,G_q$ are MV attributes. Besides to the notation  introduced in the previous section, we nees some additional notation:

\begin{enumerate}
  \item $\dot{\mathrm{\Sigma}}$ is a given set of {\em SV domain frequency constraints} -- we assume that there is exactly a domain support constraint for every SV attribute value and, then, the cardinality $\dot{m}$ of $\dot{\mathrm{\Sigma}}$ is $\dot{n}=\sum_{i=1}^{p}\dot{n}_i$; 
  \item $\ddot{\mathrm{\Sigma}}$ is a given set of {\em MV domain frequency constraints} -- we assume that there are one or more domain support constraints for every MV attribute;  for each MV attribute $G_i$,  $S_{i}$ denotes the set of itemsets of $G_i$ occurring in $\ddot{\mathrm{\Sigma}}$ ({\em frequent itemsets}), i.e., $S_{i}=$ $\{G \,|\, \langle [(G_i,a)],\sigma_1,\sigma_2\rangle \in \ddot{\mathrm{\Sigma}}\}$; the cardinality $\ddot{m}$ of $\ddot{\mathrm{\Sigma}}^1$ is $\sum_{i=1}^{q}\ddot{m}_i$, where $\ddot{m}_i= | S_{i}|$; 
  \item for each MV attribute $G_i$, $S'_{i} = \{I \subseteq \G_i \,|\ \nexists J \in S_{i}: I \subseteq J\}$ ({\em domain infrequent itemsets}) and
  $\widehat{S}'_{i}=$ $\{I \in S'_{i} \, |$ $\nexists I' \in S'_{i}: I' \subset I\}$ ({\em minimal domain infrequent itemsets}); let $\ddot{m}'_i$ be the cardinality of 
  $\widehat{S}'_{i}$ -- observe that $\ddot{m}'_i$ can be exponential in $\ddot{m}_i$ and $\ddot{n}_i$;
  \item given an infrequency (typically small) threshold $\sigma'$, $0 \leq \sigma' \leq 1$, $\ddot{\mathrm{\Sigma}}'_i(\sigma')=$ $\{\langle (G_i,I,\subseteq),0, \sigma' \rangle \, | \ I \in \widehat{S}'_i \}$ denotes the set of  {\em  infrequency constraints on the domain $\G_i$} and  $\ddot{\mathrm{\Sigma}}'(\sigma')=$ $\sum_{i=1}^{q} \ddot{\mathrm{\Sigma}}'_i(\sigma')$ is the set of the {\em domain infrequency constraints}; the cardinality $\ddot{m}'$ of $\ddot{\mathrm{\Sigma}}^1$ is $\sum_{i=1}^{q}\ddot{m}'_i$, where $\ddot{m}'_i= | \widehat{S}'_{i}| = | \ddot{\mathrm{\Sigma}}'_i(\sigma')|$; 
  \item $\widetilde{\mathrm{\Sigma}}$ is a given set of {\em  many-sorted support constraints} -- let $\widetilde{m} = |  \widetilde{\mathrm{\Sigma}}|$;
   \item $\mathrm{\Delta}$ is a given set of {\em  many-sorted duplicate constraints} -- let $m^\delta = |  \mathrm{\Delta}|$. 
\end{enumerate}

\begin{definition}\label{def:pr}
{\em 
Given $R$, $\mathrm{\dot{\Sigma}, \ddot{\Sigma}, \widetilde{\Sigma}, \Delta}$, and  two integers $\sigma' \geq 0$ and $\size > 0$,  the {\em multi-sorted inverse frequent itemset mining problem}, shortly denoted as
$\msIFM$, consists of finding a many-sorted dataset $\D$ on $R$ such that both $|\D| = \size$ and $\D \models$ 
$\mathrm{\dot{\Sigma}, \ddot{\Sigma}, \widetilde{\Sigma}, \Delta,  \ddot{\Sigma}'}(\sigma')$
(or of eventually stating that there is no such a dataset).
}
\hfill $\Box$
\end{definition}

The next result shows that $\msIFM$ reduces to some classical $\IFM$ problems if $p=0$ and $q=1$, i.e., there exists exactly one attribute in $R$ and this attribute is MV.

\begin{proposition} 
Let $P$ be the class of $\msIFM$ instances for which $p=0$, $q=1$, $\widetilde{m}=0$ and $m^\delta=0$.
Then
\begin{itemize} 
\item $P$ coincides with $\IFMI$;
\item the subclass of $P$ for which $\sigma' = \infty$ (i.e., there non infrequency constraints) coincides with $\IFM$.
\hfill $\Box$

\end{itemize}

\end{proposition}

We explicit the condition of $\msIFM$ definition as:
\allowdisplaybreaks{
\begin{align}
   \forall \langle [(A_i,a)],\sigma_1,\sigma_2\rangle \in \mathrm{\dot{\Sigma}}: \sigma_1 \leq \ \ \ \ \ \  \sum_{\mathclap{I \in \D \wedge \sat([(A_i,a)],I)}} \ \ \  \ \delta^{\D}(I) & \leq \sigma_2 \label{pc1}\\
   \forall \langle [(G_i,J,\subseteq)],\sigma_1,\sigma_2\rangle \in \mathrm{\ddot{\Sigma}}: \sigma_1 \leq \ \ \ \ \ \  \sum_{\mathclap{I \in \D \wedge \sat([(G_i,J,\subseteq)],I)}} \ \ \  \ \delta^{\D}(I) & \leq \sigma_2 \label{pc1b}\\
   \forall \langle [(G_i,J,\subseteq)],0,\delta' \rangle \in \ddot{\mathrm{\Sigma}}'(\sigma'): \ \ \ \ \ \ \ \ \sum_{\mathclap{I \in \D \wedge \sat([(G_j,J,\subseteq)],I)}} \ \ \  \delta^{\D}(I)& \leq \sigma' \label{pc2}\\
    \forall \langle L, \sigma_1, \sigma_2 \rangle \in \widetilde{\mathrm\Sigma} : \sigma_1 \leq  \ \ \ \sum_{\mathclap{I \in \D \wedge \sat(L,I)}} \ \ \  \delta^{\D}(I)& \leq \sigma_2 \label{pc3}\\
    \forall \langle L, \delta_2 \rangle \in \mathrm{\Delta}, \forall I \in \D \ s.t. \ \sat(L,I) : \ \ \  \delta^{\D}(I)& \leq \delta_2 \label{pc4} \\
 %  \forall I\in T : \delta^{\D}(I) & \leq \delta' \label{pc4}\\
%   \forall I\in \breve{S}' : \delta^{\D}(I) &=0  \label{pc4}\\
   |\D| &= \size. \:\: \ \ \ \ \label{pc5}
\end{align}
}

Note that, as the constraints (\ref{pc2}) and (\ref{pc4}) are expressed in an intensional format, they are not  explicitly given in the input. Then, the problem input size is  $\xi=\mathcal{O}( \dot{n} + \sum_{i=1}^{p} \ddot{m}_i \times \ddot{n}_i + (p+\ddot{n}) \times ( \widetilde{m}+m^\delta))$, where $p$ is the number of SV attributes, $\ddot{n}$ is the total number of items in the group attributes and 
$m_{\Sigma_D}$, $m_{\Sigma}$ and $m_{\Delta}$ are the cardinalities of $\Sigma_D$, $\Sigma$ and $\Delta$ respectively. 

To solve the problem,  we shall instantiate all the constraints (\ref{pc2}). To this end, we shall assume that the cardinality of each $\hat{S}'_{i}$ is polynomial in $\ddot{n}_i$ and $\ddot{m}'_i$ so that the total number $\hat{m}'$ of constraints (\ref{pc2}) is polynomial in the input as well. As illustrated in \cite{MocciaEtAl12,SaccaSerra12}, this is the case in practical situations and, in addition, there is a sufficient condition testing in polynomial time that 
the size of $ B_{S'_{G_j}}$ is indeed polynomial.
On the other hand, we shall leave constraints (\ref{pc4}) in an intensional format as their number could be exponential in practice - for instance, we can easily enforce that all $2^{\ddot{n}} \cdot \prod_{i=1}^{p} \dot{n}_i$ possible transactions have a given bound $\delta_2$ on their number of duplicates by simply selecting a SV attribute, say $A_1$, and by introducing the following $\dot{n}_1$ duplicate constraints: $\langle [(A_1, a_1)],\delta_2\rangle, \dots, \langle [(A_1, a_{\dot{n}_1})],\delta_2\rangle$.

\begin{proposition}\label{prop:mscomplex} \ 
\begin{enumerate}
\item Decision $\msIFM$ is $\nexp$-complete.
\item If for each $i$, $1 \leq i \leq q$, $\hat{S}'_{i}$ is polynomial in $\ddot{n}_i$ and $\ddot{m}'_i$ then
the decision version of $\msIFM$ is in $\pspace$ and $\PP$-hard.
\hfill $\Box$
\end{enumerate}
\end{proposition}

To further reduce the complexity, we relax the integer constraint for the number $\delta^\delta(I)$ of duplicates for a transaction I of a  database $\D$, i.e., 
$\delta^\delta(I)$  may be a rational number. We therefore have a relaxed version of $\msIFM$.

\section{Formulation of Relaxed ms-IFM by Succinct Linear Programming with Bounds}\label{sub:ILP}

Let $A_1, \dots,  A_p$ be the SV attributes and $G_1,\dots,G_q$ be the MV attributes.
We have that $ \A_1, \dots,  \A_p, \G_1, \dots,  \G_q$ are their  domains such that for each $i, 1 \leq i \leq p$, $|\A_i|=  \dot{n}_i$ and for each $j, 1 \leq j \leq q$, $|\G_j|=  \ddot{n}_i$.
Without loss of generality, we select any ordering of the items of each domain; in addition, we induce an ordering of the itemsets for the MV domains.

For each  $i, 1 \leq i \leq p$, we use the vector $\dot{\ti}_i=[1,\dots,\dot{n}_i]$ to list the indices of all items in $\A_i$. The index vectors for MV domains are more elaborated: for each  $i, 1 \leq i \leq q$, we use the vector $\ddot{\ti}_j=[1,\dots,2^{\ddot{n}_i}]$ to list the indices of all itemsets in $\G_i$.
In the following, to simplify the notation, we blur the difference between index and value of an item or itemset, whenever no confusion arises.

Let $x$ be a multi-dimensional array of  non-negative rational variables $x_{\dot{k}_1\, \dots\, \dot{k}_p\,\ddot{k}_1\,\dots\,\ddot{k}_q}$, where for each $i$, $1 \leq i \leq p$, 
$\dot{k}_i \in \dot{\ti}_i$ and for each $i$, $1 \leq i \leq q$, $\ddot{k}_i \in \ddot{\ti}_i$.
Given the indices $\dot{k}_1, \dots, \dot{k}_p,\ddot{k}_1,\dots,\ddot{k}_q$,  the variable $x_{\dot{k}_1\, \dots\, \dot{k}_p\,\ddot{k}_1\,\dots\,\ddot{k}_q}$, denotes the number of duplicates for the
transaction $I$ for which $\sat(L,I)$ is true, where
$$L=[(A_1,\dot{k}_1), \dots,(A_p,\dot{k}_p), (G_1,\ddot{k}_1, =), \dots,(G_1,\ddot{k}_q, =)].$$ 
\noindent
and the indices in $L$ represent the corresponding domain values.

The number of variables is $\vec{n} = 2^{\ddot{n}} \cdot \prod_{i=1}^{p} \dot{n}_i$.

For each $i$, $1 \leq i \leq \dot{n}_i$, let the vectors $s_i=[i_1,\dots,i_{\ddot{m}_i}]$ and $\hat{s}'_i=[i_1,\dots,i_{\ddot{m}'_i}]$ represent the indices of the itemsets in $S_i$ and in $\hat{S}'_i$ respectively.

% i.e.,  $S=\{I_{i_1},\dots,I_{i_m}\}$  and $B_{S'}=\{I_{j_1},\dots,I_{j_{m'}}\}$. Finally, the vector $s'$ represent all indices of the itemsets in $S'$. Note that, while $s$ is explicitly represented,   $s'$ is not: it is sufficient to check wether an index is in $s'$ and this can be easily done in linear time.

For each $\langle [(A_i, a_j)] , \sigma_1, \sigma_2 \rangle$ in $\dot{\mathrm{\Sigma}}$, where $i$ is a SV attribute index  and $j$ is a domain value index, let $\dot{l}_{i j}$ and $\dot{u}_{i j}$ denote
$\sigma_1$ and $\sigma_2$. In a similar way, for each $\langle [(G_i, I_j, \subseteq)] , \sigma_1, \sigma_2 \rangle$ in $\ddot{\mathrm{\Sigma}}$, $\ddot{l}_{i j}$ and $\ddot{u}_{i j}$ denote $\sigma_1$ and $\sigma_2$, where $j$ is the itemset index.  

Consider now $\widetilde{\mathrm{\Sigma}}$. We assume some ordering of the $\widetilde{m}$ many-sorted support constraints in it.
For each $i$, $1 \leq i \leq \widetilde{m}$, let $\widetilde{\mathrm{\Sigma}}_i= \langle L , \sigma_1, \sigma_2 \rangle$.
We denote $\sigma_1$ and $\sigma_2$ by $\tilde{l}_i$ and $\tilde{u}_i$ respectively.
In addition, $\dot{k}(\widetilde{\mathrm{\Sigma}}_i)$ denotes the list of SV attribute indices occurring in $L$ and 
$\ddot{k}_=(\widetilde{\mathrm{\Sigma}}_i)$ denotes the list of MV attribute indices occurring in an equality MV selection of $L$.

Consider now $\mathrm{\Delta}$. We assume some ordering of the $m^\delta$ many-sorted duplicate constraints in it.
For each $i$, $1 \leq i \leq m^\delta$, let $\mathrm{\Delta}_i= \langle L ,  \delta_2 \rangle$.
We denote $L$ by $L^\delta_i$ and $\delta_2$ by $u_i^\delta$.

We finally introduce a vector $w$ of $2 m +1$ non-negative rational number artificial variables, whose values represent the costs of violating some support constraints.
In particular, $w_1, \dots, w_m$ and $w_{m+1}, \dots, w_{2 m}$ are the costs of  violating respectively lower-bound and upper-bound support constraints on the itemsets in $S$ and $w_{2  m +1}$ is the cost of violating  the database size constraint.

We are now ready to formulate an approximate version of $\IFMG$ using the following linear program, whose objective function measures the cost of violating the  constraints corresponding to the artificial variables:

\small
 \begin{eqnarray}
  \LP: \ \ \text{min}  \sum_{i=1}^{p}\sum_{j=1}^{\dot{n}_i}(\dot{w}^l_{i j}+\dot{w}^u_{i j}) + \sum_{i=1}^{q}\sum_{j=1}^{\ddot{n}_i}(\ddot{w}^l_{i j}+\ddot{w}^u_{i j}) + \sum_{i=1}^{\widetilde{m}}(\widetilde{w}^l_i+\widetilde{w}^u_i)+ w^s & - \ \mathrm{s.t.} &  \label{LPof}
   \end{eqnarray}
%   \begin{eqnarray}

% \begin{align}
%w_i+ \sum_{j\in \ti} a_{ij}x_j  \geq  l_i \ \  \  i \in [1,m]     \\  w_{m+i} -\sum_{j\in \ti} a_{ij}x_j \geq  -u_i  \ i \in [1,m] \ (7)  \\ 
%-\sum_{j\in \ti}a_{i j} x_j   \geq  -\sigma' \   i \in [m+1, m'']    \\   $x_j  \leq  \delta' \ \ \   j\in \tilde{s}'   \\ 
%w_{2 m+1} +  \sum_{j\in \ti} x_j   \geq  \size$   \\ - \sum_{j\in \ti} x_j  \geq  -\size 
%%$x_j  \leq  \delta' \ \ \   j\in \tilde{s}'  $ & $w_i, x_j   \in  \mathbb{Q}^+  \ \ \  1 \leq i \leq 2m+1, j\in \ti $
%\end{align}

\allowdisplaybreaks{
   \begin{align}
    \dot{w}^l_{i j}+ \sum_{\mathclap{\dot{k} \setminus \dot{k}_i \: \ddot{k}}} \ \ x_{\dot{k}_1 \dots\, \dot{k}_{i-1}\, j\, \dot{k}_{i+1} \dots\, \dot{k}_p\, \ddot{k}_1 \dots\, \ddot{k}_q} & \geq  \dot{l}_{i j} & 1 \leq i \leq p, j \in \dot{v}_i  \label{LPl:sv}\\
    \dot{w}^u_{i j}- \sum_{\mathclap{\dot{k} \setminus \dot{k}_i \: \ddot{k}}} \ \ x_{\dot{k}_1 \dots\, \dot{k}_{i-1}\, j\, \dot{k}_{i+1} \dots\, \dot{k}_p\, \ddot{k}_1 \dots\, \ddot{k}_q} & \geq  -\dot{u}_{i j} & 1 \leq i \leq p, j \in \dot{v}_i \label{LPu:sv}\\
    \ddot{w}^l_{i j}+ \sum_{\mathclap{\dot{k}  \: \ddot{k}}} \ \ a^i_{\ddot{k}_i j} \cdot x_{\dot{k}_1 \dots\, \dot{k}_p\, \ddot{k}_1\, \dots\, \ddot{k}_{i}\, \dots\, \ddot{k}_q} & \geq  \ddot{l}_{i j} & 1 \leq i \leq q, j \in s_i  \label{LPl:mv}\\
    \ddot{w}^u_{i j}- \sum_{\mathclap{\dot{k}  \: \ddot{k}}} \ \ a^i_{\ddot{k}_i j}\cdot x_{\dot{k}_1 \dots\, \dot{k}_p\, \ddot{k}_1\, \dots\, \ddot{k}_{i}\, \dots\, \ddot{k}_q} & \geq  - \ddot{u}_{i j} & 1 \leq i \leq q, j \in s_i  \label{LPu:mv}\\
    - \sum_{\mathclap{\dot{k}  \: \ddot{k}}} \ \ a^i_{\ddot{k}_i j}\cdot x_{\dot{k}_1 \dots\, \dot{k}_p\, \ddot{k}_1\, \dots\, \ddot{k}_{i}\, \dots\, \ddot{k}_q} & \geq  - \sigma' & 1 \leq i \leq q, j \in \hat{s}'_i  \label{LP:inf}\\
   \widetilde{w}^l_{i}+ \sum_{\mathclap{\dot{k} \setminus \dot{k}(\gamma) \: \ddot{k} \setminus \ddot{k}(\gamma)}} \ \ \tilde{a}^{\ddot{k}'(\gamma)}_{\ddot{k}_{\ddot{k}'(\gamma)} \ddot{v}'(\gamma)} \cdot x_{\dot{k}(\dot{k}(\gamma) / \dot{v}(\gamma) )\, \ddot{k}((\ddot{k}(\gamma) / \ddot{v}(\gamma) ))} & \geq  \tilde{l}_{i} & 1 \leq i \leq \widetilde{m}, \gamma=\widetilde{\mathrm{\Sigma}}_i  \label{LPl:ms}\\
   \widetilde{w}^u_{i}- \sum_{\mathclap{\dot{k} \setminus \dot{k}(\gamma) \: \ddot{k} \setminus \ddot{k}(\gamma)}} \ \ \tilde{a}^{\ddot{k}'(\gamma)}_{\ddot{k}_{\ddot{k}'(\gamma)} \ddot{v}'(\gamma)} \cdot x_{\dot{k}(\dot{k}(\gamma) / \dot{v}(\gamma) )\, \ddot{k}((\ddot{k}(\gamma) / \ddot{v}(\gamma) ))} & \geq  -\tilde{u}_{i} & 1 \leq i \leq \widetilde{m}, \gamma=\widetilde{\mathrm{\Sigma}}_i  \label{LPu:ms}\\
    w^s +  \sum_{\mathclap{\dot{k}  \: \ddot{k}}} \ \ x_{\dot{k}_1 \dots\, \dot{k}_p\, \ddot{k}_1\, \dots\,  \ddot{k}_q}   &  \geq  \size &   \label{LPdim1}\\
     - \sum_{\mathclap{\dot{k}  \: \ddot{k}}} \ \ x_{\dot{k}_1 \dots\, \dot{k}_p\, \ddot{k}_1\, \dots\,  \ddot{k}_q}& \geq  -\size &   \label{LPdim2}\\
     \forall \, \dot{k}\: \ddot{k}\; \mathrm{s.t.} \: \sat(\gamma, \dot{k}\: \ddot{k}): \ \ x_{\dot{k} \ddot{k}} & \leq  u^\delta_i &  1 \leq i \leq m^\delta,  \gamma=\mathrm{\Delta}_i   \label{LPdelta}
   \end{align}
   }
   \normalsize
%%    \end{eqnarray}
%

%\vspace{0.25cm}

The variables in $w$ and in $x$  are constrained to be  non-negative rational numbers. The variables in $w$ are \emph{artificial} in the sense that their role is to absorb possible
violations of all the  constraints execpt  (\ref{LP:inf}) and (\ref{LPdim2}): the minimization of their values entails the search for a solution with the minimal number of violations. 
Therefore, the optimal solution of
the presented $\LP$ consists of a database (as induced by variables $x$ in the optimal solution) with minimal violation of the lower-bound database size
constraint.
Note that, as we do not insert artificial variables  in the Constraints (\ref{LTu}), (\ref{LPdim2}) and (\ref{LPdelta}), such constraints must be directly satisfied in any feasible solution. 
This is always possible as an initial feasible solution can be constructed as follows: $w_{2 m+1} = size_1$,
$ w_i=l_i$, $w_{m + i}=0$ and $x_j=0$ ($1 \leq i \leq m$ and $\forall j\in \ti$).

Notice that if the optimal solution of $\LP$ problem is greater than zero, then the database $\D$ induced by the optimal solution is not feasible (i.e., it is an approximate solution) for one (or both) of the following reasons: the support of at least one itemset in $S$  is not in the prescribed range or  the database size does not satisfy its lower bound.  

We use a succinct format to represent \LP.  In particular, we
simply store all the items, all the itemsets in $S$ and in $B_{S'}$, suitably represented as list of items, the vector $s$ containing the indices of the itemsets in $S$ and $B_{S'}$, the vectors $l$ and $u$ of support bounds, the database size and the values of $\sigma'$ and $\delta'$. It turns out that the input is represented in a succinct format with size $(n+n (m+m') + 2 m + 3) \cdot \omega$,
where $\omega$ is the number of bits that are used to represents constants. 
The coefficients $a_{i j}$ as well the bound constraints (\ref{LPdelta}) are computed as they are needed. We stress that the advantage of succinctness is lost unless we devise mechanisms avoiding the whole input expansion, as shown in the next sub-section.

\section{Column Generation Algorithm for Solving Relaxed ms-IFM}\label{sect:cg:simplex}
{\em Column generation} (see e.g. \cite{gilmore} and \cite{colg05})  is an extension of the simplex method for dealing with linear programs with a large number of variables.
This method solves a linear program without explicitly including all columns (i.e., variables), in the coefficient matrix but only a subset of them with cardinality equal to the number of rows (i.e., constraints).
Columns are dynamically generated by solving an auxiliary optimization problem
called the {\em pricing problem}. 

In this sub-section we extend the classical column generation simplex to handle the bounds introduced by Constraints (\ref{LPdelta}). We stress that the number of in equations implementing such constraints is exponential and, therefore, an extension of column generation is needed to handle them without expanding their representation.

The linear program to be solved is denoted as the \textit{master problem} (MP). In our case the MP problem consists of $r= 2 m + m'+2$ rows  and $c=2^n + 2 m$ columns. In addition, the variables $x_j$ with $j \in  \tilde{s}'$ are bounded by $\delta'$.

The linear program with only a subset of the MP columns with cardinality $c'$ equal to the number $r$ of rows is called the
\textit{restricted master problem} (RMP). As $r$ is polynomial in the succinct size of the input,  RMP does not need a succinct representation. Actually the number of columns $c'$ passed to RMP can be greater than $r$, provided that $c'$ is polynomial in $r$.
From linear programming theory we know that if there is an optimal solution then there also
exists an optimal solution corresponding to a basis of the coefficient matrix (in our case any basis consists of at most $r$ columns).

The column generation method looks for an optimal basis as within the simplex algorithm.
It starts from an initial basis and moves from a current basis to a new one
by replacing one basic column  with a new one with
a negative reduced cost ({\em iteration step}).
Primal feasibility is maintained and the objective function is non-increasing during this search.
The reduced cost of a column can be computed by using the current dual variables.
The task of providing a column with a negative reduced cost, or certifying that there is not such a column, is delegated to the pricing problem.
If there is no column with a negative reduced cost,
then the algorithm terminates and the current basis is optimal.

We generalize the column generation method to handle bounds as follows. Following the approach described in \cite{luenberger}, we adopt  an extended notion of basic solution, to avoid to include the bounds as constraints of the program. 
An {\em extended basic solution} is a basic feasible solution where the $n$ variables are partitioned into three groups: the set $B$ of the  classic basic variables, the set $U$ of the variables equal to the upper bound and the set $N$ of those equal to 0. 

The pseudo-code of the column generation algorithm for solving a column-succinct LP is presented in Figure \ref{AlgCGU}.
The algorithm starts by initializing $B$, that is the list of variables to be given as input to the method RMP at the first call. 
$B$ includes the indices in $w$ (i.e., the columns corresponding to the $2 m+m' + 1$ artificial variables) and those in ${s}$ (i.e., the columns corresponding to the itemsets in $S$ and in $B_{S'}$); so, the cardinality $c'$ of $B$ is equal to $r= 3 m+m'  + 1$. As discussed in the previous sub-section, a feasible solution can be easily found using such columns. The list $U$ of the variables equal to the upper bound is initially set to be empty.

The output of  RMP is: $B'$ (the list of variables in the computed basis), $Z$ (the list of values for the basic variables), $D$ (the values of the $r$ dual variables) and the updated list $U$.

Procedure  PRICE solves the pricing problem. The classical sufficient optimality condition  must be now restated for the case of \LP\ with bounds. To this end, the input of PRICE is not only the dual costs $D$ but also the list $U$, in order to exclude the itemsets in $U$ in the search of the column with the minimum reduced cost. 

PRICE returns $(j,\widetilde{c}_j)$, where $j$ is a column with a minimum reduced cost $\widetilde{c}_j$.  If $\widetilde{c}_j$ happens not to be negative, the current basis is optimal and the "while" cycle stops; otherwise, the column $j$ and all the columns in $w$ are added to the previous basis $B'$ to update $B$ and the cycle continues. Note that adding the columns in $w$ would not be necessary; nevertheless, as the number $2 m +1$ of artificial variables in $w$  is linearly bounded by the number of rows $r$, we also include all of them in $B$ to simplify the formulation of the pricing problem. The implementation of PRICE is presented in the next subsection.

\begin{algorithm}[h!]
 \SetAlgoNoLine
  \KwIn{Succinct representation of \LP, a time limit $TL$.}
  \KwOut{$B$ (list of variables in the solution basis), $Z$ (list of values for the basis variables), $U$ (list of saturated variables). } 
  	{\bf Algorithm:}\\
       Initialize $B$ = $w \cup s$, $U$ = $[]$   and STOP = \textit{false}\;
       \While{ $($ not {\em STOP and the time limit $TL$ has not  been  reached)} }{
            $(B,Z,D,U)$ = RMP$(B,U)$\;
            $(j, \widetilde{c}_j) :=$ PRICE$(D,B,U)$\;
            \uIf{$($ $\widetilde{c}_j < 0$ $)$ }
                    {$B=B \cup w \cup \{j\}$\;}
                    \uElse{STOP = \textit{true}\;}
%                    }
      }
       \textbf{return} $(B,Z)$\;
\caption{Column Generation Simplex to solve \IFMG} \label{AlgCGU}
\end{algorithm}

As the execution time could be expensive, we fix a time-limit {\em TL} for termination. The algorithm stops for one of the the following two conditions:  (i) the time-limit has been reached and (ii)  the pricing algorithm does not return a column with negative reduced cost.
The latter condition indicates that the current solution is optimal whereas in the first case, the algorithm returns a suboptimal solution.
The overall algorithm eventually terminates, provided that certain precautions against cycling are taken.  We point out that the hardest task is the implementation of procedure PRICE, that is in general  $\NP$-hard. 
Despite its alleged intractability,  the column generation algorithm has an attractive characteristic: it makes a bounded use of the space, proportional to the number $r$ of constraints and of the size of the list $U$. 

\subsection{Resolution of the Pricing Problem}

We are given a set $D$ of dual variable rational number values. We represent them by the $m$-element vectors $\lambda$ and $\pi$, the $m'$-element vector $\xi$, and the scalars $\tau_{1},$ and $\tau_{2}$ of the  RMP, that are associated to the constraints (6), (7), (8),  (10) and (11) respectively. (see Section~\ref{sub:ILP}).
Given a column $j \in v$ corresponding to any itemset variable $x_j$, the reduced cost $ \widetilde{c}_j$ is:
\begin{eqnarray*}
    \widetilde{c}_j&=&0-(\tau_{1} -\tau_{2} +\sum_{\mathclap{1 \leq i \leq m}} a_{ij}\lambda_i-\sum_{\mathclap{1 \leq i \leq m}} a_{ij}\pi_i  -\sum_{\mathclap{m+1 \leq i \leq m+m'}} a_{ij}\xi_{i-m})\\
    &=& -\tau_{1} + \tau_{2} +\sum_{\mathclap{1 \leq i \leq m}} a_{ij}(\pi_i-\lambda_i)+\ \ \sum_{\mathclap{m+1 \leq i \leq m+m'}} \ a_{ij}\xi_{i-m}.\nonumber
\end{eqnarray*}
For notational simplicity, we define $\tau = \tau_{1} -\tau_{2}$ and $\phi$ as the $(m+m')$-element vector: $$\phi=\begin{array}{|c|}
                                                                            \pi -\lambda \\
                                                                            \xi \\
                                                                          \end{array}.$$ 

Then, as  $a_{ij}=1$ if  $I_{s_i} \subseteq I_j$ or $a_{ij}=0$ otherwise, where the itemsets $I_{s_i}$ and $I_j$ correspond respectively to the row $i$ and the column $j$, the reduced cost can be reformulated as:
$$\widetilde{c}_j= -\tau+\sum_{1 \leq i \leq m+m',\, I_{s_i} \subseteq I_j}\phi_i.\nonumber $$

We formulate the Pricing Problem in terms of an integer linear program  that computes
an itemset $I^*$, say with index $j$, such that $j \in v$ and $\widetilde{c}_j$ is minimum.
$I^*$ is represented by a vector of binary variables $\beta=[\beta_1,\dots,\beta_n]$, corresponding to the $n$ items: each component $\beta_h$  indicates whether $I^*$ contains the item $o_h$ ($\beta_h=1$) or not ($\beta_h=0$).
We use the vector of binary variables $y=[y_1,\dots,y_{m+m'}]$, corresponding to the itemsets in $s$, to model the inclusion of such itemsets in $I^*$:
thus,  $y_{i} = 1$ if $I_{s_i} \subseteq I^*$ or $y_{i} = 0$ otherwise. Then $I^*$ is the union of all itemsets $I_{s_i}\in S$  for which $y_{i} = 1$. Note that, since the price problem may not generate a column already present in $U$, also the set $U$ plays an important role in the price formulation. Thus,  for each column $j \in U$, we define a value $k_j$ as the number of itemsets $I \in S$ contained in $I_j$, that is $$k_j=|\{i| \leq i \leq m+m', I_{s_i}\subset I_j\}|$$.

The integer linear program formulation to solve the pricing problem, denoted as \texttt{PRICE}, follows (to simplify the notation we set $m''=m+m'$):

%\begin{scriptsize}  
\begin{align}
\texttt{PRICE}: \ \text{minimize} \ \ \sum_{\mathclap{1 \leq i \leq m''}} \phi_i y_{i}  & & \label{ob}
\end{align}
\begin{align}
  \sum_{\mathclap{o_h\in I_{s_i}}} \beta_h + 1  \leq |I_{s_i}| +y_i  & & 1 \leq i \leq m'' \label{c1}  \\
\beta_h  - \ \ \ \sum_{\mathclap{1 \leq i \leq m'',\, o_h\in I_{s_i}}} \  y_i \leq 0    & &  1 \leq h \leq n  \label{c5} \\
 y_i  \leq  \beta_h  & & \begin{array}{c}1 \leq i \leq m'' \\ \forall o_h\in I_{s_i}\end{array} \label{c2}\\
\sum_{\mathclap{\begin{tiny}\begin{array}{c}1 \leq i \leq m'',\\I_{s_i}\subset I_j\end{array} \end{tiny}}}\ y_i- k_j \   \sum_{\mathclap{\begin{tiny}\begin{array}{c}1 \leq i \leq m'',\\ I_{s_i}\not \subset I_j\end{array}\end{tiny}}}y_i  \leq  k_j-1 & &\forall j\in U\\
0\, \leq\,  y_i \leq 1 & & 1 \leq i \leq m''  \label{c3} \\
 \beta_h  \in \{0,1\}& & 1 \leq h \leq n . \label{c4}
\end{align}
%\end{scriptsize}

The objective function (\ref{ob}) represents the reduced cost $\widetilde{c}_j$ (modulo the constant $-\tau$) of a generic column $j \in v$, that we want to minimize to compute $I^*$. 

The constraints (\ref{c1}) impose that, given any $i$,  if  $\beta_h=1$ for all $o_h\in I_{s_i}$ then  $y_i=1$; in other words, $I^*$ contains all items of an itemset $I_{s_i}$.
The constraints (\ref{c5}) impose that if $\beta_h=1$ then there exists an element $i$ of $s$ such that $o_h \in I_{s_i} \wedge y_i=1$; thus, an item $o_h$ is in $I^*$ only if at least one of the itemsets included in $I^*$ contains $o_h$.
The constraints (\ref{c2}) impose that if $\exists\: o_h\in I_{s_i}\::\beta_h=0$ then  $y_i$ must be equal to zero, i.e., an itemset $I_{s_i}$ cannot be declared included in $I^*$ if any of its items is not contained in $I^*$.
The constraints (\ref{c2}) also enforce that, if $I_{s_i}$ is declared to be an itemset included in $I^*$ ($y_i = 1$), then all items of it must be in $I^*$ as well ($\beta_h= 1, \forall h \in I_{s_i}$).
The constraint (\ref{c0}), as proven in Proposition~\ref{pco}, imposes that no column $j \in U$ is returned from pricing method.

Observe that it is not necessary to explicitly enforce integer constraints  on the variables $y$ in constraints (\ref{c3}). In fact, for each $i$-element of $s$ two cases are possible: (i) each variable $\beta_h$ with $o_h\in I_{s_i}$ has value $1$, or (ii) there exists at least one variable $\beta_h$ with $o_h\in I_{s_i}$ that has value $0$. In the first case, by constraints (\ref{c1}) $y_i = 1$. Instead in the second case,  $y_i =0$ by constraints (\ref{c2}). Hence, as the $\beta$ variables are enforced to be integer, variables $y$ can only be either 1 or 0. 

The crucial point of the ILP formulation is that it excludes the columns in $U$ in the search of the column with minimum reduced cost. This is done because of the following result.

\begin{proposition}\label{pco}
Each column $j \in U$ is an infeasible solution for the PLI price formulation. 
\hfill $\Box$
\end{proposition}

\section{Conclusion}\label{sect:concl}

In this paper we have presented an extension of the  inverse frequent set mining problem (\IFM), called $\msIFM$, in order to generate big table instances that reflect given frequency and infrequency patterns. 
We have assumed that the scheme of a big table is of the form $R(K, A_1, \dots,  A_p, G_1,\dots,$ $G_q )$, where $K$ is the table key, $A_1, \dots,  A_p$ are SV attributes and
$G_1,\dots,G_q$ are MV attributes. The frequency and infrequency patterns are of three types:
domain support constraint, multi-sorted supported constraints and multi-sorted duplicate constraints.
The $\msIFM$ has been formulated as a succinct linear program and and extension of the column-generation simplex has been adopted to solve the linear program by the invention of a suitable solution for the pricing problem.

In our setting,  we do not have defined lower bound duplicate constraints for many-sorted transactions. On going research is devoted to handle lower bound as well.To this end, we intend to use some technicalities, e.g., a lower bound constraint $\delta \leq x_{j^1 \dots j^p j}$ can be simply implemented by replacing 
it with $0 \leq x'_{j^1 \dots j^p j}$, where $x'_{j^1 \dots j^p j} = x_{j^1 \dots j^p j} - \delta$. Obviously support constraints must be accordingly rewritten.

%\vspace{-0.3cm}
\bibliographystyle{abbrv}
\bibliography{cbiblio}

\end{document}